\documentclass[pra,aps,amsmath,amssymb,eqsecnum,showkeys]{revtex4}

\newcommand{\pen}{\openone}
\newcommand{\hh}{{\mathcal{H}}}

\newcommand{\lnp}{\mathcal{L}}
\newcommand{\lsa}{\mathcal{L}_{s.a.}}
\newcommand{\lsp}{\mathcal{L}_{+}}
\newcommand{\tr}{{\mathrm{tr}}}
\newcommand{\id}{{\mathrm{id}}}
\newcommand{\spp}{{\mathrm{supp}}}
\newcommand{\cmb}{{\mathbb{B}}}
\newcommand{\mpb}{{\mathbb{P}}}

\newcommand{\cla}{{\mathcal{A}}}
\newcommand{\clb}{{\mathcal{B}}}
\newcommand{\mc}{{\mathcal{M}}}
\newcommand{\nc}{{\mathcal{N}}}
\newcommand{\ppc}{{\mathcal{P}}}
\newcommand{\qpc}{{\mathcal{Q}}}
\newcommand{\bro}{\boldsymbol{\rho}}
\newcommand{\bsg}{\boldsymbol{\sigma}}
\newcommand{\nil}{\boldsymbol{0}}
\newcommand{\am}{{\mathsf{A}}}
\newcommand{\gms}{{\mathsf{G}}}
\newcommand{\km}{{\mathsf{K}}}
\newcommand{\mm}{{\mathsf{M}}}
\newcommand{\nm}{{\mathsf{N}}}
\newcommand{\ppm}{{\mathsf{P}}}
\newcommand{\qpm}{{\mathsf{Q}}}
\newcommand{\ax}{{\mathsf{X}}}
\newcommand{\ay}{{\mathsf{Y}}}

\newcommand{\rmd}{{\mathrm{D}}}

\begin{document}
\clearpage
\preprint{}

\title{On the Brukner--Zeilinger approach to information in quantum measurements}

\author{Alexey E. Rastegin}
\affiliation{Department of Theoretical Physics, Irkutsk State University,
Gagarin Bv. 20, Irkutsk 664003, Russia}

\begin{abstract}
We address the problem of properly quantifying information in
quantum theory. Brukner and Zeilinger proposed the concept of an
operationally invariant measure based on measurement statistics.
Their measure of information is calculated with probabilities
generated in a complete set of mutually complementary
observations. This approach was later criticized for several
reasons. We show that some critical points can be overcome by
means of natural extension or reformulation of the
Brukner--Zeilinger approach. In particular, this approach is
connected with symmetric informationally complete measurements.
The ``total information'' of Brukner and Zeilinger can further be
treated in the context of mutually unbiased measurements as well
as general symmetric informationally complete measurements. The
Brukner--Zeilinger measure of information is also examined in the
case of detection inefficiencies. It is shown to be decreasing
under the action of bistochastic maps. The Brukner--Zeilinger
total information can be used for estimating the map norm of
quantum operations.
\end{abstract}


\maketitle

\pagenumbering{arabic}
\setcounter{page}{1}

\section{Introduction}\label{sc1}

Quantum information science has currently made impressive advances
in both theory and practice \cite{bh2013}. Feynman emphasized that
quantum systems are very hard to be simulated at the classical
level \cite{feynman82}. On the other hand, such negative claim
also inspires a positive reason for trying to build quantum
computers \cite{feynman86}. Quantum key distribution has provided
a long-term technological solution already implemented in a lot of
commercial products \cite{assche06,pivk10}. Quantum algorithms
allow to solve efficiently a number of important problems, which
are currently intractable \cite{bacon10,nori14}. Developments in
quantum information processing stimulated a renewed interest to
foundations of quantum mechanics. This subject is a thriving,
lively and controversial field of research \cite{az02,briggs13}.
Currently, conceptual questions are often reformulated in
information-theoretic terms. Actually, results of a quantum
measurement are finally recorded in some row of statistical data.
Hence, we have come across a problem to quantify an amount of
information that could be extracted from such data.

The problem of determining quantum state quite differs from the
classical formulation. There are many possible scenarios to be
imagined. Attacking a system of quantum key distribution, Eve is
typically bused with discriminating between two or more
alternatives known to her {\it a priori}. During an individual
attack, she captures only a single information carrier. An
opposite situation deals with a very large ensemble of identical
copies. In practice, a number of copies is never infinite though
large. Our experience leads to the following conclusion. The
proportion of times that the given outcome occurs settles down to
some value as number of trials becomes larger and larger. The
ultimate value of this proportion is meant as the probability of
the given outcome. Dealing with quantum systems, the observer can
take different experiments, which might even completely exclude
each other. For example, the state of a spin-$1/2$ system is often
considered to be estimated with measurements of the three
orthogonal components of spin \cite{brz99}. In more than two
dimensions, such complementary measurements are formulated in
terms of the so-called mutually unbiased bases (MUBs). This
concept was actually considered by Schwinger \cite{schwinger}.

To approach properly an informational measure, Brukner and
Zeilinger considered the following situation \cite{brz99}. Suppose
that we know probabilities of all outcomes and try to guess a
number of occurrences of the prescribed outcome among finite
experimental trials. Of course, our prediction will allow an
amount of uncertainty, which can be estimated with respect to some
confidence interval. Taking an uncertainty per single trial and
summing it for all outcomes, one naturally leads to a measure of
uncertainty in one experiment \cite{brz99}. It is shown to be
$1$ minus the sum of squared probabilities. Hence, Brukner and
Zeilinger defined a measure of information in one experiment and
in a set experiments. For $d+1$ MUBs, the corresponding total
information turned to be operationally invariant in the following
sense \cite{brz99}. The sum of the individual measures of
information for mutually complementary observations is invariant
with respect to a choice of the particular set of complementary
observations. In other words, this sum is invariant under unitary
rotations of the measured state. The latter implies that there is
no information flow between the system of interest and its
environment \cite{brz99}.

Mutually unbiased bases are an interesting mathematical object as
well as an important tool in many physical issues \cite{bz10}.
Such bases can be used in quantum key distribution, state
reconstruction, quantum error correction, detection of quantum
entanglement, and other topics. Mutually unbiased bases are
connected with symmetric informationally complete
measurements. A positive operator-valued measure (POVM) is said to
be informationally complete, if its statistics determine
completely the quantum state \cite{busch91,dps04}. To increase an
efficiency at determining the state, elements of such a
measurement should have rank one. An informationally complete POVM
is called symmetric, when all pairwise inner products between the
POVM elements are equal \cite{rbsc04}. In general, the maximal
number of MUBs in $d$ dimensions is still an open question
\cite{bz10}. When $d$ is a prime power, the answer $d+1$ is known
\cite{bz10}. Constructions of $d+1$ MUBs for such $d$ rely on
properties of prime powers and on an underlying finite field
\cite{wf89}. It also seems to be hard to get a unified way for
building a symmetric informationally complete POVM (SIC-POVM) in
all dimensions.

The authors of \cite{kag14} introduced the concept of mutually
unbiased measurements. The core idea is that elements of such a
measurement are not rank one. This method does not reach the
maximal efficiency but is easy to construct. It turns out that a
complete set of $d+1$ mutually unbiased measurements can be built
explicitly for arbitrary finite $d$ \cite{kag14}. An utility of
such measurements in quantum information science deserves further
investigations. It is also unknown whether rank-one SIC-POVMs
exist in all finite dimensions. The positive answer was obtained
with a weaker condition that POVM elements are not rank one. The
authors of \cite{kgour13} proved the existence of general
SIC-POVMs in all finite dimensions. It is not insignificant that
general SIC-POVMs can be constructed within a unified approach.
Studies of mutually unbiased measurements and general SIC-POVMs
were continued in \cite{fei14,rastpsic,fei15,rastosid}. We will
show that these measurements are interesting in the context of the
Brukner--Zeilinger approach \cite{brz99,brz01,brz02,brz09}. This
approach to quantifying an amount of information will be shown to
be realizable within three additional types of quantum
measurements.

The paper is organized as follows. In Section \ref{sec2},
preliminary material is reviewed. In particular, we recall the
definitions of mutually unbiased measurements and general
SIC-POVMs. Section \ref{sec3} is devoted to a general discussion
of the Brukner--Zeilinger approach to quantification of
information in quantum measurements. Its treatment in terms of
Tsallis' entropies of degree $2$ is mentioned. In Section
\ref{sec4}, we show that an operationally invariant measure of
information can be approached within the three measurement
schemes. They are respectively based on a single SIC-POVM, on a
set of $d+1$ mutually unbiased measurements, and on a general
SIC-POVM. These measurement schemes give an alternative to $d+1$
MUBs known only for prime power dimensions. In Section \ref{sec5},
the Brukner--Zeilinger approach is examined for the case of
detection inefficiencies, when the ``no-click'' events are
allowed. In Section \ref{sec6}, we show that the
Brukner--Zeilinger total information cannot increase under the
action of bistochastic maps. Relations between the
Brukner--Zeilinger approach and non-unitality are examined in Section
\ref{sec7}. In Section \ref{sec8}, we conclude the paper with a
summary of results.

\section{Preliminaries}\label{sec2}

In this section, we review the required material on mutually
unbiased measurements and general SIC-POVMs. Let $\lnp(\hh_{d})$
be the space of linear operators on $d$-dimensional Hilbert space
$\hh_{d}$. By $\lsp(\hh_{d})$, we denote the set of positive
semidefinite operators on $\hh_{d}$. By $\lsa(\hh_{d})$, we mean
the $d^{2}$-dimensional real space of Hermitian operators on
$\hh_{d}$. A state of $d$-level system is represented by density
operator $\bro\in\lsp(\hh_{d})$ normalized as $\tr(\bro)=1$. For
operators $\ax,\ay\in\lnp(\hh_{d})$, their Hilbert--Schmidt inner
product is defined by \cite{watrous1}
\begin{equation}
\langle\ax{\,},\ay\rangle_{\mathrm{HS}}:=\tr(\ax^{\dagger}\ay)
\ . \label{hsdef}
\end{equation}
Quantum measurements are commonly dealt in terms of the POVM
formalism \cite{peresq}. We consider a set of elements
$\am_{j}\in\lsp(\hh_{d})$ such that the completeness relation
holds, namely
\begin{equation}
\sum\nolimits_{j} \am_{j}=\pen_{d}
\ . \label{cmprl}
\end{equation}
Here, the $\pen_{d}$ denotes the identity operator on $\hh_{d}$.
The set $\cla=\{\am_{j}\}$ is called a (POVM). For pre-measurement state $\bro$, the probability
of $j$-th outcome is written as \cite{peresq}
\begin{equation}
p_{j}(\cla|\bro)=\tr(\am_{j}\bro)
\ . \label{njpr}
\end{equation}
It is of key importance that the number of different outcomes can
be more than the dimensionality of $\hh_{d}$ \cite{peresq}. Of
course, in practice POVM measurements involve auxiliary systems,
so that degrees of freedom are actually added.

Let $\clb^{(1)}=\bigl\{|b_{j}^{(1)}\rangle\bigr\}$ and
$\clb^{(2)}=\bigl\{|b_{k}^{(2)}\rangle\bigr\}$ be two orthonormal
bases in $\hh_{d}$. They are mutually unbiased if and only if for
all $j$ and $k$,
\begin{equation}
\bigl|\langle{b}_{j}^{(1)}|b_{k}^{(2)}\rangle\bigr|=\frac{1}{\sqrt{d}}
\ . \label{twb}
\end{equation}
The set $\cmb=\bigl\{\clb^{(1)},\ldots,\clb^{(L)}\bigr\}$ is
formed by mutually unbiased bases (MUBs), when each two bases from
this set are mutually unbiased. The measurement in one basis cannot
give anything about the state, which was prepared
in another basis. This property is essential in some schemes of
quantum key distribution.

Let us recall symmetric informationally complete
measurements. In $d$-dimensional Hilbert space, we consider a set
of $d^{2}$ rank-one operators of the form
\begin{equation}
\nm_{j}=\frac{1}{d}
\>|\phi_{j}\rangle\langle\phi_{j}|
\ . \label{usic}
\end{equation}
If the normalized vectors $|\phi_{j}\rangle$ all satisfy the condition
\begin{equation}
\bigl|\langle\phi_{j}|\phi_{k}\rangle\bigr|^{2}=\frac{1}{d+1}
\qquad (j\neq{k})
\ , \label{undn1}
\end{equation}
the set $\nc=\{\nm_{j}\}$ is a symmetric informationally complete
POVM (SIC-POVM) \cite{rbsc04}. It was conjectured that SIC-POVMs
exist in all dimensions \cite{appl2005}. The existence of
SIC-POVMs has been shown analytically or numerically for all
dimensions up to 67 \cite{grassl10}. For a discussion of
connections between MUBs and SIC-POVMs, see \cite{ruskai09} and
references therein. Weyl--Heisenberg (WH) covariant SIC-sets of
states in prime dimensions are examined in \cite{adf07}. WH
SIC-sets, whenever they exist, consist solely of minimum
uncertainty states with respect to R\'{e}nyi's $2$-entropy for a
complete set of MUBs \cite{adf07}. The authors of \cite{dabo2014}
derived bounds on accessible information and informational power
for the case of SIC-sets of quantum states. In general,
informationally complete sets of positive matrices are discussed
in the book \cite{hiai2014}. The authors of \cite{krsw05}
discussed approximate versions of a SIC-POVM, when a small
deviation from uniformity of the inner products is allowed.

Basic constructions of MUBs concern the case, when $d$ is a prime
power. If $d$ is another composite number, maximal sets of MUBs
are an open problem \cite{bz10}. We can try to approach
``unbiasedness'' with weaker conditions. The authors of
\cite{kag14} proposed the concept of mutually unbiased
measurements. They consider two POVM measurements
$\ppc=\{\ppm_{j}\}$ and $\qpc=\{\qpm_{k}\}$. Each of them contains
$d$ elements such that
\begin{align}
& \tr(\ppm_{j})=\tr(\qpm_{k})=1
\ , \label{tmn1}\\
& \tr(\ppm_{j}\qpm_{k})=\frac{1}{d}
\ . \label{dmn1}
\end{align}
Thus, the POVM elements are all of trace one, but now not of rank
one. The formula (\ref{dmn1}) replaces (\ref{twb}). The
Hilbert--Schmidt product of two elements from the same POVM
depends on the so-called efficiency parameter $\varkappa$
\cite{kag14}. It holds that
\begin{equation}
\tr(\ppm_{j}\ppm_{k})=\delta_{jk}{\,}\varkappa
+(1-\delta_{jk}){\>}\frac{1-\varkappa}{d-1}
\ , \label{mjmk}
\end{equation}
and similarly for the elements of $\qpc$. The efficiency parameter
obeys \cite{kag14}
\begin{equation}
\frac{1}{d}<\varkappa\leq1
\ . \label{vklm}
\end{equation}
For $\varkappa=1/d$ we have the trivial case, in which
$\ppm_{j}=\pen_{d}/d$ for all $j$. The value $\varkappa=1$, when
possible, leads to the standard case of mutually unbiased bases.
More precise bounds on $\varkappa$ will depend on a construction
of measurement operators. The efficiency parameter shows how close
the measurement operators are to rank-one projectors \cite{kag14}.
For the given $\varkappa$, we take the set
$\mpb=\bigl\{\ppc^{(1)},\ldots,\ppc^{(L)}\bigr\}$ of POVMs
satisfying (\ref{mjmk}). When each two POVMs also obey conditions
of the forms (\ref{tmn1}) and (\ref{dmn1}), the set $\mpb$ is a
set of mutually unbiased measurements (MUMs). Allowing
$\varkappa\neq1$, the authors of \cite{kag14} built $d+1$ MUMs in
$d$-dimensional Hilbert space for arbitrary $d$. Their
construction is based on the generators of ${\textup{SU}}(d)$. For
the given $d$, the parameter $\varkappa$ ranges in the interval,
which is determined by the smallest or largest eigenvalues of some
traceless operators. In this regard, we cannot fix $\varkappa$
without specifying $d$. Of course, the efficiency parameter should
approach $1$ as close as possible.

Similar ideas can be used in building general SIC-POVMs. For all
finite $d$, a common construction has been given \cite{kgour13}.
Consider a POVM with $d^{2}$ elements $\mm_{j}$, which satisfy the
following two conditions. First, for all $j=1,\ldots,d^{2}$ one
has
\begin{equation}
\tr(\mm_{j}\mm_{j})=a
\ . \label{ficnd}
\end{equation}
Second, the pairwise inner products are all symmetrical, namely
\begin{equation}
\tr(\mm_{j}\mm_{k})=b
\qquad (j\neq{k})
\ . \label{secnd}
\end{equation}
Then the operators $\mm_{j}$ form a general SIC-POVM. Combining
the conditions (\ref{ficnd}) and (\ref{secnd}) with the
completeness relation finally gives \cite{kgour13}
\begin{equation}
b=\frac{1-ad}{d(d^{2}-1)}
\ . \label{bvia}
\end{equation}
We also get $\tr(\mm_{j})=1/d$ for all $j=1,\ldots,d^{2}$.
Therefore, the value $a$ is the only parameter that characterizes
the type of a general SIC-POVM. This parameter is restricted as
\cite{kgour13}
\begin{equation}
\frac{1}{d^{3}}<{a}\leq\frac{1}{d^{2}}
\ . \label{resa}
\end{equation}
The value $a=1/d^{3}$ corresponds to the case
$\mm_{j}=\pen_{d}/d^{2}$, which does not give an informationally
complete POVM. The value $a=1/d^{2}$ is achieved, when the POVM
elements are all rank-one \cite{kgour13}. The latter is actually
the case of usual SIC-POVMs, when POVM elements are represented in
terms of the corresponding unit vectors as (\ref{usic}). Even if
SIC-POVMs exist in all dimensions, they are rather hard to
construct. Similarly to usual SIC-POVMs, general SIC-POVMs have a
specific structure that makes them appropriate in determining an
informational content of a quantum state.

In Section \ref{sec5}, we will use monotonicity of the relative
entropy under the action of trace-preserving completely positive
(TPCP) maps. So, we recall some required material. Let us consider
a linear map
\begin{equation}
\Phi:{\>}\lnp(\hh_{d})\rightarrow\lnp(\hh_{m})
\ . \label{phhp}
\end{equation}
To describe physical processes, linear maps have to be completely
positive \cite{nielsen,bengtsson}. Let $\id_{n}$ be the identity
map on $\lnp(\hh_{n})$, where the $n$-dimensional space $\hh_{n}$
is assigned to a reference system. The complete positivity implies
that the map $\Phi\otimes\id_{n}$ is positive for all $n$.
Completely positive maps are often called quantum operations
\cite{nielsen}. Each completely positive map can be represented in
the form \cite{watrous1,nielsen}
\begin{equation}
\Phi(\ax)=\sum\nolimits_{i}\km_{i}{\,}\ax{\,}\km_{i}^{\dagger}
\ . \label{osrp}
\end{equation}
Here, the Kraus operators $\km_{i}$ map the input space $\hh_{d}$
to the output space $\hh_{m}$. The map preserves the trace, when
the Kraus operators satisfy
\begin{equation}
\sum\nolimits_{i}\km_{i}^{\dagger}{\,}\km_{i}=\pen_{d}
\ . \label{clrl}
\end{equation}
Trace-preserving quantum operations are usually referred to as
quantum channels. Applying to a POVM measurement, the formula
(\ref{clrl}) merely gives the completeness relation
\cite{nielsen}.

\section{On definition of the Brukner--Zeilinger information}\label{sec3}

Quantum theory can shortly be characterized as a formal scheme for
representing states together with rules for computing the
probabilities of different outcomes of an experiment
\cite{peresq}. In this regard, the notion of quantum state is
rather a list of the statistical properties of an ensemble of
identically prepared systems. In a series of papers
\cite{brz99,brz01,brz02,brz09}, Brukner and Zeilinger considered
the question of informational content of an unknown quantum state.
To quantify the amount of information, a prospective measure
should have some natural properties. These properties are also
connected with a proper choice of individual experiments or rather
a set of experiments. Choosing experiments, the observer can
actually manage different kinds of information that will manifest
themselves, although the total amount of information is apparently
limited \cite{az02}.

Let us consider an experiment, in which a non-degenerate
$d$-dimensional observable is measured. This test is actually
connected with the corresponding basis
$\clb=\bigl\{|b_{j}\rangle\bigr\}$. As a rule, the observer has
only a limited number of systems to work with. Keeping the
probability distribution
$p_{j}(\clb|\bro)=\langle{b}_{j}|\bro|b_{j}\rangle$, the observer
try to guess how many times a specific outcome will occur. In such
situation, the number of occurrences of some outcome in future
repetitions cannot be expected precisely \cite{brz99}. The authors
of \cite{brz99} suggested to characterize the experimenter's
uncertainty by the quantity
\begin{equation}
U_{BZ}(\clb|\bro):=1-\sum\nolimits_{j=1}^{d} p_{j}(\clb|\bro)^{2}
\ . \label{ubzdf}
\end{equation}
This approach is motivated with considering mean-square-deviation
for uncertainty in the number of occurrences. It will be
convenient to introduce the index of coincidence
\begin{equation}
C(\clb|\bro):=\sum\nolimits_{j=1}^{d} p_{j}(\clb|\bro)^{2}
\ . \label{icdf}
\end{equation}
We then have $U_{BZ}(\clb|\bro)=1-C(\clb|\bro)$. The case of
complete lack of information in an experiment corresponds to the
uniform distribution. Hence, Brukner and Zeilinger proposed to
define the measure of information as \cite{brz99,brz01}
\begin{equation}
I_{BZ}(\clb|\bro):=\sum_{j=1}^{d} \left(p_{j}(\clb|\bro)-\frac{1}{d}\right)^{\!2}
. \label{ibzdf}
\end{equation}
In principle, the right-hand side of (\ref{ibzdf}) could be
rescaled by appropriate normalization factor \cite{brz99}. The
latter is chosen with respect to the context. Since the uniform
distribution is obtained with the completely mixed state
$\bro_{*}=\pen_{d}/d$, we can rewrite the Brukner--Zeilinger
information as
\begin{equation}
I_{BZ}(\clb|\bro)=C(\clb|\bro)-C(\clb|\bro_{*})
\ . \label{ibzcc}
\end{equation}
As we will see, this form is useful in studies of the case with
detection inefficiencies. Here, the uniform distribution is not a
good reference point for comparison. It is also convenient for
generalizing the approach to POVM measurements. Indeed, for a POVM
measurement the number of outcomes typically exceeds
dimensionality \cite{peresq}.

When the observer have many copies of the same quantum state, he
will rather tend to measure the state in several mutually
complementary bases. For example, the state of spin-$1/2$ could be
measured along one of three orthogonal axes. The authors of
\cite{brz99,brz01} defined the total information content by
summarizing the measures (\ref{ibzdf}) for all complementary
tests. Suppose that we have the set $\cmb$ of $d+1$ MUBs in
$d$-dimensional space. For any density matrix
$\bro\in\lsp(\hh_{d})$, one then gives \cite{larsen90,ivan92}
\begin{equation}
\sum_{\clb\in\cmb}C(\clb|\bro)=1+\tr(\bro^{2})
\ . \label{dp1b}
\end{equation}
Thus, the sum of indices of coincidence is determined by the
quantity $\tr(\bro^{2})$ usually called purity \cite{bengtsson}.
Then the total information is represented as
\begin{equation}
\sum_{\clb\in\cmb}I_{BZ}(\clb|\bro)=\tr(\bro^{2})-\tr(\bro_{*}^{2})
=\tr(\bro^{2})-\frac{1}{d}
\ . \label{intotb}
\end{equation}
It must be stressed here that this quantity is invariant under
unitary transformations of $\bro$. When we have a set
$\cmb_{L}=\bigl\{\clb^{(1)},\ldots,\clb^{(L)}\bigr\}$ of $L$ MUBs,
there holds \cite{molm09}
\begin{equation}
\sum_{\clb\in\cmb_{L}}C(\clb|\bro)\leq\frac{L-1}{d}+\tr(\bro^{2})
\ . \label{mmb}
\end{equation}
For the case $L<d+1$, the sum of $L$ indices of coincidence
cannot be determined in terms of purity solely. Hence, we can only
write the inequality
\begin{equation}
\sum_{\clb\in\cmb_{L}}I_{BZ}(\clb|\bro)\leq\tr(\bro^{2})-\tr(\bro_{*}^{2})
\ . \label{mintotb}
\end{equation}
The left-hand side of (\ref{mintotb}) is generally changed under
unitary transformations of $\bro$.

The question about invariance or non-invariance under unitary
transformations can be illustrated with the three
spin-1/2 measurements along mutually orthogonal axes \cite{brz01}.
For one and the same spin state, the three coordinate axes may be
oriented arbitrarily. Here, the total information (\ref{intotb})
does not depend on such an orientation. Indeed, any axes rotation
can be reformulated as a unitary transformation of the given
state. The eigenbases of the three Pauli observables are mutually
unbiased, whence the total information (\ref{intotb}) is invariant
under unitary transformations of the state.

The Shannon entropy is one of the basic notions of
information theory. If a measurement is described by the
probabilities $p_{j}(\clb|\bro)$, then the Shannon entropy is
written as
\begin{equation}
H_{1}(\clb|\bro):=-\sum\nolimits_{j=1}^{d} p_{j}(\clb|\bro){\>}\ln{p}_{j}(\clb|\bro)
\ . \label{shdf}
\end{equation}
Summing the Shannon measures for all the bases, we obtain some
total characteristic. It turned out that such total characteristic
is generally not invariant under unitary transformations. The
authors of \cite{brz01} clearly exemplified this fact with the
three spin-$1/2$ measurements along orthogonal axes. As a total
measure of informational character, the sum of three Shannon
entropies has several counter-intuitive properties \cite{brz01}.
First, it can be different for states of the same purity. Second,
it changes in time even for a completely isolated system. Third,
it depends on particular details of an experimental setup. Even in
two dimensions, therefore, the mentioned approach to quantifying
information in quantum measurements seems to be inappropriate.

Thus, the sum of the Shannon entropies of generated probability
distributions is generally not invariant even for the case, when
$d+1$ MUBs exist. In opposite, the total information
(\ref{intotb}) is constant here. Note that the Brukner--Zeilinger
information can be interpreted in entropic terms. For
$0<\alpha\neq1$, the Tsallis $\alpha$-entropy of generated
probability distribution $p_{j}(\clb|\bro)$ is defined by
\begin{equation}
H_{\alpha}(\clb|\bro):=\frac{1}{1-\alpha}{\>}
\biggl(\,\sum_{j=1}^{d}{p_{j}(\clb|\bro)^{\alpha}} -1\biggr)
\, . \label{tsaldf}
\end{equation}
This entropy is widely used in non-extensive statistical mechanics
due to Tsallis \cite{tsallis}. For $\alpha=2$, the corresponding
Tsallis entropy is connected with the index of coincidence as
\begin{equation}
H_{2}(\clb|\bro)=1-C(\clb|\bro)
\ . \label{h2ic}
\end{equation}
Hence, we represent the Brukner--Zeilinger information as
\begin{equation}
I_{BZ}(\clb|\bro)=H_{2}(\clb|\bro_{*})-H_{2}(\clb|\bro)
\ . \label{ibzh2}
\end{equation}
Thus, the Brukner--Zeilinger measure shows a reduction in the
uncertainty due to a deviation of the density matrix from the
completely mixed one. However, the uncertainty is quantified by
the Tsallis entropy of degree $\alpha=2$.

\section{Three schemes with special types of quantum measurements}\label{sec4}

In this section, we  will discuss use of the Brukner--Zeilinger
approach with a SIC-POVM, with a complete set of MUMs, and with a
general SIC-POVM. In each of these cases, we finally obtain an
information measure operationally invariant in the terminology of
\cite{brz99}. To apply the result (\ref{dp1b}), we have to perform
$d+1$ projective measurements, if the required MUBs all exist. So,
it is interesting to examine the Brukner--Zeilinger total
information with other quantum measurements. For a POVM
$\cla=\{\am_{j}\}$, we define
\begin{equation}
I_{BZ}(\cla|\bro)=C(\cla|\bro)-C(\cla|\bro_{*})
\ , \label{imzcc}
\end{equation}
where $C(\cla|\bro)$ is the sum of all squared probabilities of
the form (\ref{njpr}). The definition (\ref{imzcc}) is a natural
generalization of the formula (\ref{ibzcc}).

We first mention that a single POVM measurement is sufficient for
our purposes. Suppose that $\nc=\{\nm_{j}\}$ is a symmetric
informationally complete POVM in $d$ dimensions. As was shown in
\cite{rastepjd}, the corresponding index of coincidence is equal
to
\begin{equation}
C(\nc|\bro)=\sum\nolimits_{j=1}^{d^{2}}p_{j}(\nc|\bro)^{2}=\frac{\tr(\bro^{2})+1}{d(d+1)}
\ . \label{indc0}
\end{equation}
That is, for a SIC-POVM the index of coincidence is expressed in
terms of purity of the given density matrix. For the completely
mixed state, we have
\begin{equation}
C(\nc|\bro_{*})=\frac{\tr(\bro_{*}^{2})+1}{d(d+1)}=\frac{1}{d^{2}}
\ . \label{indc0cms}
\end{equation}
For a SIC-POVM, the Brukner--Zeilinger information is represented
as
\begin{equation}
I_{BZ}(\nc|\bro)=\frac{\tr(\bro^{2})-\tr(\bro_{*}^{2})}{d(d+1)}
\ . \label{intotn}
\end{equation}
This quantity is merely obtained by dividing the total information
(\ref{intotb}) by $d(d+1)$. In this regard, the quantity
(\ref{intotn}) can also be treated as a measure of total
information. It is important since SIC-POVMs could exist for those
values of $d$, for which $d+1$ MUBs do not exist. Say, for MUBs we
do not know the answer already for $d=6$, whereas the existence of
SIC-POVMs has been shown for $d\leq67$ \cite{grassl10}. Of course,
any SIC-POVM is more complicated for implementation than a single
projective measurement. However, we need $d+1$ projective
measurements for calculating (\ref{intotb}). Even if $d+1$ MUBs
exist, the scheme with them may require more costs than the scheme
based on a single SIC-POVM. In this respect, the result
(\ref{intotn}) is also significant. At the same time,
constructions of SIC-POVMs for sufficiently larger $d$ may rather
be complicated. We will further see that the Brukner--Zeilinger
concept of total information can be developed with $d+1$ MUMs and
with a general SIC-POVM. These types of measurement are
interesting in the sense that each of them allows a unified
theoretical description.

For arbitrary $d$, we can built a set of $d+1$ MUMs of some
efficiency $\varkappa<1$ \cite{kag14}. We shall now
consider the Brukner--Zeilinger approach with such measurements.
Let $\mpb$ be a set $d+1$ MUMs of the efficiency $\varkappa$ in
$d$-dimensional space. As was shown in \cite{rastosid,fei15}, we
then have
\begin{equation}
\sum_{\ppc\in\mpb}C(\ppc|\bro)
=1+\frac{1-\varkappa+(\varkappa{d}-1){\,}\tr(\bro^{2})}{d-1}
\ . \label{ubpd1}
\end{equation}
For pure states, the right-hand side of (\ref{ubpd1}) reads
$1+\varkappa$. The latter was obtained in \cite{kag14}. With a set
$\mpb_{L}=\bigl\{\ppc^{(1)},\ldots,\ppc^{(L)}\bigr\}$ of $L$ MUMs,
we can only write the inequality \cite{rastosid}
\begin{equation}
\sum_{\ppc\in\mpb_{L}}C(\ppc|\bro)\leq\frac{L-1}{d}+\frac{1-\varkappa+(\varkappa{d}-1){\,}\tr(\bro^{2})}{d-1}
\ . \label{ubp1}
\end{equation}
Due to (\ref{ubpd1}), we have arrived at a conclusion. For the
complete set $\mpb$ of $d+1$ MUMs of the efficiency $\varkappa$
and any density matrix $\bro\in\lsp(\hh_{d})$, one gives
\begin{equation}
\sum_{\ppc\in\mpb}I_{BZ}(\ppc|\bro)=
\frac{\varkappa{d}-1}{d-1}\bigl[\tr(\bro^{2})-\tr(\bro_{*}^{2})\bigr]
\, . \label{intotm}
\end{equation}
The right-hand side of (\ref{intotm}) increases proportionally to
the efficiency parameter $\varkappa$. At the prescribed
efficiency, the sum of the Brukner--Zeilinger information measures
is determined by purity solely. For $\varkappa=1$, the result
(\ref{intotm}) is reduced to (\ref{intotb}). The latter, however,
depends on the existence of a complete set of mutually
complementary observables. Among other critical points, this fact
was mentioned in \cite{hall00}. On the other hand, a set of $d+1$
MUMs with some $\varkappa<1$ has been constructed for arbitrary
$d$ \cite{kag14}. Except for $\varkappa=1$, mutually unbiased
measurements are not projective. Together, a set of $d+1$ MUMs
involves $d(d+1)$ POVM elements. The scheme with a general
SIC-POVM seems to be more effective, since it involves only
$d^{2}$ POVM elements.

Let us proceed to the case of general SIC-POVMs. It is
interesting, since general SIC-POVMs can be built within a scheme
common for all $d$ \cite{kgour13}. In opposite, a unified approach
to constructing SIC-POVMs with rank-one elements hardly exists.
Moreover, the existence of usual SIC-POVMs for all $d$ is
plausible but still not proved. For a general SIC-POVM $\mc$, we
have \cite{rastpsic}
\begin{equation}
C(\mc|\bro)=\sum\nolimits_{j=1}^{d^{2}}p_{j}(\mc|\bro)^{2}
=\frac{(ad^{3}-1){\,}\tr(\bro^{2})+d(1-ad)}{d(d^{2}-1)}
\ . \label{gindc0}
\end{equation}
where the parameter $a$ characterizes this POVM. Due to
(\ref{gindc0}), for any density matrix $\bro\in\lsp(\hh_{d})$ we
then get
\begin{equation}
I_{BZ}(\mc|\bro)=\frac{ad^{3}-1}{d(d^{2}-1)}\bigl[\tr(\bro^{2})-\tr(\bro_{*}^{2})\bigr]
\, . \label{gintotn}
\end{equation}
This quantity expresses the total information associated with the
general SIC-POVM $\mc$. For $a=1/d^{2}$, the result
(\ref{gintotn}) is naturally reduced to (\ref{intotn}). Thus, the
Brukner--Zeilinger approach to quantifying total information of
the given quantum state can be realized, at least in principle,
with mutually unbiased measurements as well as with a general
SIC-POVM.

In this section, we have shown that the Brukner--Zeilinger concept
of total information can be realized within the three measurement
schemes. They are respectively based on a single SIC-POVM, on a
set of $d+1$ MUMs, and on a general SIC-POVM. We are sure in
existence of the complete set of MUBs only for specific values of
the dimensionality. We can also recall that even the case $d=6$ is
still not understood. For this reason, an alternative realization
of the Brukner--Zeilinger approach is certainly interesting. On
the other hand, implementation of such experimental schemes may be
not easy due to very special structure of measurement operators.
So, the developed approach should take into account a role of
detection inefficiencies. In this regard, the authors of
\cite{safin06} criticized the Brukner--Zeilinger approach. In the
next section, we examine the question in more details.

\section{Formulation for measurements with detection inefficiencies}\label{sec5}

In practice, measurement devices inevitably suffer from losses.
The authors of \cite{safin06} considered the Brukner--Zeilinger
approach in the case of non-zero probability of the no-click
event. For definiteness, we first describe this case for
complementary measurements in MUBs. Let the parameter
$\eta\in[0;1]$ characterize a detector efficiency. The no-click
event is presented by additional outcome $\varnothing$. Assume
that for any basis $\clb$ the inefficiency-free distribution
$\bigl\{p_{j}(\clb|\bro)\bigr\}$ is altered as
\begin{equation}
p_{j}^{(\eta)}(\clb|\bro)=\eta{\,}p_{j}(\clb|\bro)
\ , \qquad
p_{\varnothing}^{(\eta)}(\clb|\bro)=1-\eta
\ . \label{dspd1}
\end{equation}
In other words, we mean detectors of the same efficiency for all
of the used MUBs. This assumption seems to be physically natural
and has been adopted in \cite{safin06}. In essence, the above
formulation coincides with the first model of detection
inefficiencies applied in \cite{rchtf12}. On the other hand, the
authors of \cite{rchtf12} focus on measurements in cycle scenarios
of the Bell type.

It was noticed that the Brukner--Zeilinger approach may have some
doubts in application to more realistic models of the experiment.
In principle, we could expect that the total information should
vanish with negligible $\eta$. At a glance, however, one comes
across an opposite situation. The authors of \cite{safin06}
illustrated this conclusion with the three spin-$1/2$
measurements along orthogonal axes. They calculated the sum
of three quantities of the form (\ref{ibzdf}) for different
$\eta\in[0;1]$ and found the following. First, the minimum of the
sum is reached at some intermediate value of $\eta>0$. Second, for
$\eta\to0^{+}$ the sum becomes even larger than for the
inefficiency-free case $\eta=1$. Such results gave a ground for
criticizing the Brukner--Zeilinger approach \cite{safin06}.

In our opinion, these doubts may be overcome with a proper
modification of the form (\ref{ibzdf}). Here, we compare obtained
probability distributions with the uniform one. However, such a
comparison is meaningful only in the inefficiency-free case
$\eta=1$. In the distribution (\ref{dspd1}), one of probabilities
depends on detectors solely. As its value is $1-\eta$, the
uniform distribution does not have actual bearing for the case
$\eta<1$. Instead, we propose to compare the actual probability
distribution with the distribution obtained with the completely
mixed input. It is quite reached by replacing (\ref{ibzdf}) with
(\ref{ibzcc}). More precisely, for the case of detection
inefficiencies we use the quantity
\begin{equation}
I_{BZ}^{(\eta)}(\clb|\bro)=C^{(\eta)}(\clb|\bro)-C^{(\eta)}(\clb|\bro_{*})
=H_{2}^{(\eta)}(\clb|\bro_{*})-H_{2}^{(\eta)}(\clb|\bro)
\ . \label{ibzccet}
\end{equation}
The superscripts emphasize here that the information measures are
all calculated with actual ``distorted'' probabilities. Apparently,
preparing the completely mixed state is not difficult. For the
existing experimental setup, therefore, statistics with the
completely mixed input can be observed and stored. Stored data
will be used in future for applications of the definition
(\ref{ibzccet}). Thus, we shall consider a more realistic case of
detection inefficiencies on the base of (\ref{ibzccet}). It was
shown in \cite{rastqqt} that for all $\alpha>0$ we have
\begin{equation}
H_{\alpha}^{(\eta)}(\clb|\bro)=\eta^{\alpha}H_{\alpha}(\clb|\bro)+h_{\alpha}(\eta)
\ , \label{qtlm0}
\end{equation}
where $H_{\alpha}^{(\eta)}(\clb|\bro)$ is the $\alpha$-entropy of
``distorted'' distribution (\ref{dspd1}). Of course, the binary
entropy $h_{\alpha}(\eta)$ is written as
\begin{equation}
h_{\alpha}(\eta)=\frac{1}{1-\alpha}\>
\bigl(\eta^{\alpha}+(1-\eta)^{\alpha}-1\bigr)
\> . \label{bnta}
\end{equation}
For $\alpha=1$, results of the form (\ref{qtlm0}) were applied in
studying entropic Bell inequalities with detector inefficiencies
\cite{rchtf12}. We will also assume that for a POVM
$\cla=\{\am_{j}\}$ the inefficiency-free probabilities
$p_{j}(\cla|\bro)$ are actually altered similarly to
(\ref{dspd1}). For $\alpha=2$, we then have
\begin{equation}
H_{2}^{(\eta)}(\cla|\bro)=\eta^{2}H_{2}(\cla|\bro)+h_{2}(\eta)
\ . \label{qtlm02}
\end{equation}
The left-hand side of (\ref{qtlm02}) is the entropy calculated
with actual measurement statistics.

We can now reformulate the results (\ref{intotb}), (\ref{intotn}),
(\ref{intotm}), (\ref{gintotn}) in the case of detection
inefficiencies. It is for this reason that we modified definition
of the Brukner--Zeilinger information according to
(\ref{ibzccet}). That is, the terms with $\bro_{*}$ also take into
account an influence of no-click events. Combining (\ref{intotb})
with (\ref{qtlm0}) for $\alpha=2$, we have arrived at a
conclusion. When $d+1$ MUBs exist and form the set $\cmb$, the
total information with actually observed statistics is equal to
\begin{equation}
\sum_{\clb\in\cmb}I_{BZ}^{(\eta)}(\clb|\bro)=\eta^{2}\bigl[\tr(\bro^{2})-\tr(\bro_{*}^{2})\bigr]
\, . \label{intotbet}
\end{equation}
When the parameter $\eta$ decreases, the total information also
decreases proportionally to the square of $\eta$. With a
negligible efficiency of detection, no information about the
system could be obtained. This very natural picture motivates the
proposed definition (\ref{ibzccet}).

Using the described model of inefficiencies, we further obtain the
following relations. If a POVM $\nc$ is symmetric informationally
complete then
\begin{equation}
I_{BZ}^{(\eta)}(\nc|\bro)=\eta^{2}{\>}\frac{\tr(\bro^{2})-\tr(\bro_{*}^{2})}{d(d+1)}
\ . \label{intotnet}
\end{equation}
This result is obtained by combining (\ref{intotn}) with
(\ref{qtlm02}). For the complete set $\mpb$ of $d+1$ MUMs of the
efficiency $\varkappa$, we also rewrite (\ref{intotm}) as
\begin{equation}
\sum_{\ppc\in\mpb}I_{BZ}^{(\eta)}(\ppc|\bro)=
\eta^{2}{\>}\frac{\varkappa{d}-1}{d-1}\bigl[\tr(\bro^{2})-\tr(\bro_{*}^{2})\bigr]
\, . \label{intotmet}
\end{equation}
Due to (\ref{gintotn}), for a general SIC-POVM $\mc$ we have
\begin{equation}
I_{BZ}^{(\eta)}(\mc|\bro)=
\eta^{2}{\>}\frac{ad^{3}-1}{d(d^{2}-1)}\bigl[\tr(\bro^{2})-\tr(\bro_{*}^{2})\bigr]
\, . \label{gintotnet}
\end{equation}
The right-hand side of any of the formulas
(\ref{intotbet})--(\ref{gintotnet}) monotonically increases with
the detection efficiency $\eta$. Thus, criticism related to
detection inefficiencies is truly overcome by a proper
modification of the definition of the Brukner--Zeilinger measure.
The idea is that the probability distribution used for referencing
should take into account the parameter $\eta$. In principle, the
results (\ref{intotbet})--(\ref{gintotnet}) could be adopted for
verification of concrete experimental setups with respect to their
efficiency. Of course, the inefficiency model used is very simple
in character. Probably, more sophisticated models of detection
inefficiencies could be developed. Nevertheless, our discussion has
shown that the Brukner--Zeilinger approach can quite be placed in
the context of real experiments with a limited efficiency.

\section{Monotonicity under the action of bistochastic maps}\label{sec6}

We have seen that, for some special measurements, the
Brukner--Zeilinger total information can exactly be expressed in
terms of purity of the quantum state of interest. In effect, the
four information measures (\ref{intotb}), (\ref{intotn}),
(\ref{intotm}), (\ref{gintotn}) are all proportional to the
quantity
\begin{equation}
\tr(\bro^{2})-\tr(\bro_{*}^{2})=\tr(\bro^{2})-\frac{1}{d}
\ . \label{trrhr}
\end{equation}
So, we can treat it as a quantum measure of informational content
of the given quantum state. The author of \cite{luo07} showed the
following fact. The quantity (\ref{trrhr}) is directly connected
with usual quantum-mechanical variance averaged over every
orthonormal basis in $\lsa(\hh_{d})$. Hence, the
Brukner--Zeilinger concept of invariant information is supported
within a more traditional point of view. We will provide another
interesting interpretation for (\ref{trrhr}). This interpretation
allows to study monotonicity of the Brukner--Zeilinger information
under the action of quantum stochastic maps.

The relative entropy is a very important measure of statistical
distinguishability \cite{nielsen}. In the classical regime, the
relative entropy is also known as the Kullback-Leibler divergence
\cite{KL51}. Its extension to entropic functions of the Tsallis
type was discussed in \cite{borland,fky04}. Let
$\spp(\bro)\subseteq\hh_{d}$ be the subspace spanned by those
eigenvectors that correspond to strictly positive eigenvalues of
$\bro$. This subspace is typically called the support of $\bro$
\cite{nielsen}. For density operators $\bro$ and $\bsg$, the
quantum relative entropy is expressed as \cite{nielsen}
\begin{equation}
\rmd_{1}(\bro||\bsg):=
\begin{cases}
\tr(\bro\,\ln\bro-\bro\,\ln\bsg) {\ },
& \text{if $\spp(\bro)\subseteq\spp(\bsg)$} {\ }, \\
+\infty{\ }, & \text{otherwise} {\ }.
\end{cases}
\label{relan}
\end{equation}
Many fundamental results of quantum information theory are closely
related to properties of the relative entropy
\cite{nielsen,vedral02}.

The divergence (\ref{relan}) was generalized in several ways. To
connect the Brukner--Zeilinger approach, we will use quantum
divergences of the Tsallis type. For $\alpha\in(1;+\infty)$, the
Tsallis $\alpha$-divergence is defined as
\begin{equation}
\rmd_{\alpha}(\bro||\bsg):=
\begin{cases}
\frac{1}{\alpha-1}\bigl[\tr(\bro^{\alpha}\bsg^{1-\alpha})-1\bigr]{\,},
& \text{if $\spp(\bro)\subseteq\spp(\bsg)$}{\ }, \\
+\infty{\ }, & \text{otherwise}{\ }.
\end{cases}
\label{qendf}
\end{equation}
For $\alpha\in(0;1)$, we merely use the first entry without
conditions. Up to a factor, this relative entropy is a particular
case of quasi-entropies introduced by Petz \cite{petz86}.
Quasi-entropies are a quantum counterpart of Csisz\'{a}r's
$f$-divergences \cite{ics67}. For more details, see the papers
\cite{ruskai10,hmpb11} and references therein. It is easy to see
that the quantity (\ref{trrhr}) can be represented as
\begin{equation}
\tr(\bro^{2})-\tr(\bro_{*}^{2})=\frac{1}{d}{\>}\rmd_{2}(\bro||\bro_{*})
\ . \label{da2rr}
\end{equation}
This formula gives a connection of the Brukner--Zeilinger total
information with the Tsallis relative entropy.

One of the basic properties of the quantum relative entropy is its
monotonicity under the action of trace-preserving completely
positive (TPCP) maps \cite{nielsen}. As has been shown, the four
information measures (\ref{intotb}), (\ref{intotn}),
(\ref{intotm}), (\ref{gintotn}) are invariant with respect to
unitary transformations. Keeping the measurement setup, we now aim
to compare the Brukner--Zeilinger measure before and after the
action of TPCP maps. For this reason, we will focus on the case of
the same input and output space. Then Kraus operators of the
operator-sum representation (\ref{osrp}) are expressed by square
matrices.

In classical regime, the relative $\alpha$-entropy of Tsallis'
type is monotone for all $\alpha\geq0$ \cite{fky04}. Due to
non-commutativity, the quantum case is more complicated in
character. The quantum $\alpha$-divergence (\ref{qendf}) is
monotone under the action of TPCP maps for $\alpha\in(0;2]$. That
is, for $\alpha\in(0;2]$ and arbitrary TPCP map $\Phi$ we have
\begin{equation}
\rmd_{\alpha}\bigl(\Phi(\bro)\big|\big|\Phi(\bsg)\bigr)\leq\rmd_{\alpha}(\bro||\bsg)
\ . \label{mnren}
\end{equation}
This claim is based on the general approach of \cite{hmpb11} and
the following results of matrix analysis. The function
$\xi\mapsto\xi^{\alpha}$ is matrix concave on $[0;+\infty)$ for
$0\leq\alpha\leq1$ and matrix convex on $[0;+\infty)$ for
$1\leq\alpha\leq2$ (see, respectively, theorems 4.2.3 and 1.5.8 in
\cite{bhatia07}).

Bistochastic maps form an important class of TPCP maps. Recall
that we consider the case of the same input and output space.
Taking arbitrary operators $\ax,\ay\in\lnp(\hh_{d})$, the adjoint
map is defined by \cite{watrous1}
\begin{equation}
\bigl\langle\Phi(\ax),\ay\bigr\rangle_{\mathrm{HS}}=
\bigl\langle\ax,\Phi^{\dagger}(\ay)\bigr\rangle_{\mathrm{HS}}
\ . \label{adjm}
\end{equation}
For the completely positive map (\ref{osrp}), its adjoint is represented as
\begin{equation}
\Phi^{\dagger}(\ax)=\sum\nolimits_{i} \km_{i}^{\dagger}{\,}\ax{\,}\km_{i}
\ . \label{aosrp}
\end{equation}
If this adjoint is trace preserving, then Kraus operators of $\Phi$
also obey
\begin{equation}
\sum\nolimits_{i}\km_{i}{\,}\km_{i}^{\dagger}=\pen_{d}
\ . \label{rlcl}
\end{equation}
If a quantum map is completely positive and its Kraus operators
satisfy both (\ref{clrl}) and (\ref{rlcl}) the map is called
bistochastic \cite{bengtsson}. Bistochastic maps can be treated as
a quantum counterpart of bistochastic matrix, which act in the
space of probability vectors. The principal fact is that the
completely mixed state is a fixed point of any bistochastic map,
namely
\begin{equation}
\Phi(\bro_{*})=\bro_{*}
\ . \label{cmsfp}
\end{equation}
This property is referred to as unitality of the map
\cite{nielsen,bhatia07}. Combining (\ref{mnren}) with
(\ref{cmsfp}), we have arrived at a conclusion. For
$\alpha\in(0;2]$ and all bistochastic maps
$\Phi:{\>}\lnp(\hh_{d})\rightarrow\lnp(\hh_{d})$, one gets
\begin{equation}
\rmd_{\alpha}\bigl(\Phi(\bro)\big|\big|\bro_{*}\bigr)\leq\rmd_{\alpha}(\bro||\bro_{*})
\ . \label{mnren2}
\end{equation}
We will use (\ref{mnren2}) for $\alpha=2$. Thus, the quantity
(\ref{da2rr}) cannot increase under the action of bistochastic
maps. In other words, for bistochastic maps we write
\begin{equation}
\tr\bigl(\Phi(\bro)^{2}\bigr)-\tr(\bro_{*}^{2})\leq\tr(\bro^{2})-\tr(\bro_{*}^{2})
\ . \label{dar2r}
\end{equation}
Due to (\ref{dar2r}), we see that the quantities (\ref{intotb}),
(\ref{intotn}), (\ref{intotm}), (\ref{gintotn}) can only decrease
under the action of bistochastic maps. As was recently shown in
\cite{bzhpl15}, bistochastic quantum operation can only increase
quantum entropies of very general class.

Since the four measures (\ref{intotb}), (\ref{intotn}),
(\ref{intotm}), (\ref{gintotn}) depend on purity of the state,
they are all invariant with respect to unitary transformation. In
the terminology of \cite{brz99}, they are all operationally
invariant measures of information. The unitary invariance has been
treated as one of basic reasons for using just this approach to
quantification of information in quantum measurements. Further,
the above information measures cannot increase under the action of
bistochastic maps. For a bistiochastic map, its adjoint is a TPCP
map as well. Here, the property (\ref{cmsfp}) plays a
key role. Quantum fluctuation theorems form another direction, in
which unitality seems to be very important. As was claimed in
\cite{albash}, unitality replaces microreversibility as the
restriction for the physicality of reverse processes.
Significance of unitality or non-unitality of quantum stochastic
maps deserves further investigations. In the next section, we will
discuss some relations between this question and the
Brukner--Zeilinger total information.

\section{Non-unital maps and the Brukner--Zeilinger approach}\label{sec7}

We have seen that the quantity (\ref{trrhr}) can only decrease
under the action of bistichastic maps. It is natural to expect
that (\ref{trrhr}) may be increased for non-unital quantum
operations. In this section, we will study connections of the
Brukner--Zeilinger total information with characterization of such
maps. The latter seems to be closely related with quantum
fluctuation theorems. Recent advances in dealing with small
quantum systems have led to growing interest in their
thermodynamics \cite{jareq11}. A certain progress has been
connected with studies of the Jarzynski equality \cite{jareq97a}
and related fluctuation theorems \cite{talhag07,cht11}. Recent
studies are mainly concentrated on formulations for open quantum
systems \cite{motas11,vedral12,kafri12,goold15,azz15}. Some of
such results have been shown to be valid in the case of
bistochastic maps \cite{albash,rast13}. Jarzynski equality for
quantum stochastic maps can naturally be formulated in terms of
the non-unitality operator \cite{arkz14}. It turns out that norms
of this operator can be evaluated within the Brukner--Zeilinger
approach.

Operators of interest are often characterized by means of norms.
Some of them are especially important. To each
$\ax\in\lnp(\hh_{d})$, we assign $|\ax|\in\lsp(\hh_{d})$ as the
unique positive square root of $\ax^{\dagger}\ax$. The eigenvalues
of $|\ax|$ counted with multiplicities are the singular values of
$\ax$ denoted by $s_{j}(\ax)$. For $q\in[1;\infty]$, the Schatten
$q$-norm is defined as \cite{watrous1}
\begin{equation}
\|\ax\|_{q}:=\Bigl(\sum\nolimits_{j=1}^{d} s_{j}(\ax)^{q}{\,}\Bigr)^{1/q}
\ . \label{schnd}
\end{equation}
This family includes the trace norm $\|\ax\|_{1}=\tr|\ax|$ for
$q=1$, the Hilbert--Schmidt norm
$\|\ax\|_{2}=\langle\ax,\ax\rangle_{\mathrm{HS}}^{1/2}$ for $q=2$,
and the spectral norm
\begin{equation}
\|\ax\|_{\infty}=\max\bigl\{s_{j}(\ax):{\>}1\leq{j}\leq{d}\bigr\}
\label{innm}
\end{equation}
for $q=\infty$. These norms are widely used in quantum information
theory. They also give a tool for characterizing linear maps. For
a linear map $\Phi$, its norm is defined as
\begin{equation}
\|\Phi\|:=\sup\bigl\{\|\Phi(\ax)\|_{\infty}:{\>}\|\ax\|_{\infty}=1\bigr\}
\>. \label{pnrm0}
\end{equation}
We will use the following fact proved, e.g., in item 2.3.8 of
\cite{bhatia07}. If a map
$\Phi:{\>}\lnp(\hh_{d})\rightarrow\lnp(\hh_{d})$ is positive, then
\begin{equation}
\|\Phi\|=\|\Phi(\pen_{d})\|_{\infty}
\ . \label{pnrm1}
\end{equation}
In terms of the completely mixed state, we write
$\|\Phi\|=d{\,}\|\Phi(\bro_{*})\|_{\infty}$.

For a linear map $\Phi:{\>}\lnp(\hh_{d})\rightarrow\lnp(\hh_{d})$,
the non-unitality operator is written as \cite{arkz14}
\begin{equation}
\gms_{\Phi}:=\Phi(\bro_{*})-\bro_{*}
\ . \label{htxd}
\end{equation}
This operator is zero for all bistochastic maps. For TPCP maps,
the Hilbert--Schmidt norm of (\ref{htxd}) is immediately expressed
in terms of the Brukner--Zeilinger measure of information. Indeed,
the squared Hilbert--Schmidt norm of $\gms_{\Phi}$ is written as
\begin{equation}
\bigl\langle\Phi(\pen_{d})-\pen_{d},\Phi(\pen_{d})-\pen_{d}\bigr\rangle_{\mathrm{HS}}=
\tr\bigr(\Phi(\pen_{d})^{2}\bigr)-2\,\tr\bigr(\Phi(\pen_{d})\bigr)+d=
\tr\bigr(\Phi(\pen_{d})^{2}\bigr)-d
\ . \label{fsnrm}
\end{equation}
Here, we recall that $\Phi(\pen_{d})\in\lsp(\hh_{d})$ is Hermitian
and $\tr\bigr(\Phi(\pen_{d})\bigr)=d$ due to preservation of the
trace. Dividing (\ref{fsnrm}) by $d^{2}$ and taking the square
root, for a TPCP map we have
\begin{equation}
\|\gms_{\Phi}\|_{2}=\sqrt{\tr\bigr(\Phi(\bro_{*})^{2}\bigr)-1/d}
=\sqrt{\tr\bigr(\Phi(\bro_{*})^{2}\bigr)-\tr(\bro_{*}^{2})}
\ . \label{hsnnuo}
\end{equation}
Thus, obtaining the Brukner--Zeilinger total information allows
also to calculate the Hilbert--Schmidt norm of the non-unitality
operator.

The difference
$\tr\bigr(\Phi(\bro_{*})^{2}\bigr)-\tr(\bro_{*}^{2})$ can be
evaluated by means of measurements schemes described in Sections
\ref{sec3} and \ref{sec4}. When an unknown quantum channel is
given as some black box, we prepare the completely mixed state
with putting it into the black box. The output $\Phi(\bro_{*})$ is
further subjected to one of measurement schemes available for the
given $d$. This run is repeated as many times as required for
collecting measurement statistics. Statistical data should be
sufficient for evaluation of the left-hand side of one of the
relations (\ref{intotb}), (\ref{intotn}), (\ref{intotm}), and
(\ref{gintotn}). Thus, we obtain the quantity (\ref{trrhr}) for
$\bro=\Phi(\bro_{*})$ and apply (\ref{hsnnuo}).

Using the result (\ref{hsnnuo}), for quantum operations we can
estimate from above the map norm (\ref{pnrm1}). We will use a
relation between vector norms proved in \cite{rastepjd}. It was
later applied for deriving fine-grained uncertainty relations for
a set of MUBs and a set of MUMs \cite{rastqip15}. As follows from
the results of appendix A of \cite{rastepjd}, for any operator
$\ax\in\lnp(\hh_{d})$ we have
\begin{equation}
\|\ax\|_{\infty}\leq
\frac{1}{d}\left(\|\ax\|_{1}+\sqrt{d-1}\,\sqrt{d\,\|\ax\|_{2}^{2}-\|\ax\|_{1}^{2}}\right)
. \label{appax}
\end{equation}
For a TPCP map, we have
$\|\Phi(\bro_{*})\|_{1}=\tr\bigl(\Phi(\bro_{*})\bigr)=1$.
Combining this with (\ref{appax}) gives
\begin{equation}
\|\Phi(\bro_{*})\|_{\infty}\leq
\frac{1}{d}\left(1+\sqrt{d-1}\,\sqrt{d\,\|\Phi(\bro_{*})\|_{2}^{2}-1}\right)
. \label{appaph}
\end{equation}
Due to (\ref{hsnnuo}) and
$\|\Phi(\bro_{*})\|_{2}^{2}=\tr\bigl(\Phi(\bro_{*})^{2}\bigr)$,
multiplying (\ref{appaph}) by $d$ leads to
\begin{equation}
\|\Phi\|\leq
1+\sqrt{d(d-1)}\,\|\gms_{\Phi}\|_{2}
\ . \label{nmphi}
\end{equation}
Thus, for quantum operations the map norm (\ref{pnrm0}) is bounded
from above in terms of the Hilbert--Schmidt norm of the
corresponding non-unitality operator. For bistochastic maps, we
have $\|\Phi\|=\|\pen_{d}\|_{\infty}=1$ and $\gms_{\Phi}=\nil$, so
that the inequality (\ref{nmphi}) is saturated here.

The above findings can further be illustrated with the following
example. Let $\bigl\{|i\rangle\bigr\}_{i=1}^{d}$ be an orthonormal
basis in $\hh_{d}$. We consider the quantum operation
$\Psi:{\>}\lnp(\hh_{d})\rightarrow\lnp(\hh_{d})$ with Kraus
operators
\begin{equation}
\km_{i}=|i_{0}\rangle\langle{i}|
\ , \label{kii0}
\end{equation}
where $|i_{0}\rangle$ is some prescribed state of the basis. This
map represents the complete contraction to a pure state. Taking
$|i_{0}\rangle$ as a ground state, one can describe the process of
spontaneous emission in atomic physics. In a certain sense,
quantum operations of the form (\ref{kii0}) enjoy extreme
non-unitality. The condition (\ref{clrl}) is clearly satisfied,
whereas
\begin{equation}
\Psi(\pen_{d})=\sum\nolimits_{i=1}^{d}\km_{i}{\,}\km_{i}^{\dagger}
=d\,|i_{0}\rangle\langle{i}_{0}|
\ . \label{0kii}
\end{equation}
In this example, we have
$\Psi(\bro_{*})=|i_{0}\rangle\langle{i}_{0}|$ and
$\tr\bigr(\Psi(\bro_{*})^{2}\bigr)-\tr(\bro_{*}^{2})=1-1/d$.
Hence, the Brukner--Zeilinger information reaches
its maximal value. We also note that the inequality
(\ref{nmphi}) is saturated here. Indeed, substituting the term
$\|\gms_{\Psi}\|_{2}=\sqrt{1-1/d}$ into the right-hand side of
(\ref{nmphi}) results in the value $d$ that is exactly
$\|\Psi\|=\|\Psi(\pen_{d})\|_{\infty}$. Thus, quantum operations
of the form (\ref{kii0}) show a behavior quite opposite to
bistochastic maps.

\section{Conclusion}\label{sec8}

We have considered the Brukner--Zeilinger approach to quantifying
information in quantum measurements on a finite-level system. This
problem is essential due to recent advances in quantum information
processing. The original formulation of Brukner and Zeilinger was
based on projective measurements in the complete set of MUBs. This
formulation is therefore restricted, since even the case of MUBs
in dimensionality $6$ is still not resolved \cite{bz10}. We have
shown that the idea of operationally invariant measure of
information can truly be realized within the three schemes based
on special types of quantum measurements. Namely, these schemes
respectively use a single SIC-POVM, a complete set of MUMs, and a
single general SIC-POVM. Such measurements are easy to construct.
In addition, costs on the schemes with a single SIC-POVM may be
less. The Brukner--Zeilinger measure of information was also
criticized on the following ground. In real experiments, the
``no-click'' events inevitably occur. Some doubts in the case of
detection inefficiencies were discussed in \cite{safin06}. Such
criticism is overcome by means of natural reformulation of the
approach considered. Namely, the uniform distribution is a good
reference only for the inefficiency-free case. Otherwise, we
should use for comparison some probability distribution that takes
into account a real efficiency of detectors. The desired
probability distribution is naturally obtained by putting the
completely mixed state into real experiments. The corresponding
data can be stored and further used for calculating required
quantities. Information measures of the Brukner--Zeilinger type
are not only unitarily invariant, they cannot also increase under
the action of bistochastic maps. Using this approach for
characterization of non-unital TPCP maps is considered. If a
quantum channel is given as black box, the measurement schemes
described can be used for determining the Hilbert--Schmidt norm of
the non-unitality operator. Potential applications of information
measures of the Brukner--Zeilinger type in quantum information
science deserve further investigations. The authors of
\cite{dm2015} recently proposed the constructor theory of
information, which is aimed to derive the properties of
information entirely from the laws of physics. It would be
interesting to study measures of information in quantum theory
within  the constructor theory.

\acknowledgments

I am grateful to anonymous reviewers for useful comments.

\end{document}